\documentclass{mem}
\usepackage{natbib}\usepackage{txfonts}\usepackage{balance}
\usepackage{graphicx}
\usepackage[a4paper,breaklinks,dvipdfm]{hyperref}
\idline{75}{282}
\begin{document}
\def\teff{$T\rm_{eff }$}
\def\kms{$\mathrm {km s}^{-1}$}

\title{
Intermediate-mass black holes in Globular Clusters
}

   \subtitle{}

\author{
N. \,L\"utzgendorf \inst{1} 
          \and
          M. Kissler-Patig\inst{2}
          \and
          K. Gebhardt\inst{3}          
          \and
          H. Baumgardt\inst{4}          
          \and
          E. Noyola\inst{5,6}
		  \and
          P. T. de Zeeuw\inst{1,7}
          \and
          N. Neumayer\inst{1}
          \and
          B. Jalali\inst{8}
          \and
          A. Feldmeier \inst{1}
          }


   \institute{European Southern Observatory (ESO),
              Karl-Schwarzschild-Strasse 2, 85748 Garching, Germany\\
              \email{nluetzge@eso.org}
         \and     
              Gemini Observatory, Northern Operations Center, 670 N. A'ohoku Place
Hilo, Hawaii, 96720, USA
         \and
			 Astronomy Department, University of Texas at Austin, 
			 Austin, TX 78712, USA 
         \and
			 School of Mathematics and Physics, University of Queensland, 
			 Brisbane, QLD 4072, Australia
         \and
             Instituto de Astronomia, Universidad Nacional Autonoma de Mexico (UNAM), 
             A.P. 70-264, 04510 Mexico
         \and
			 University Observatory, Ludwig Maximilians University, 
			 81679 Munich, Germany	 
         \and
			 Sterrewacht Leiden, Leiden University, 
			 Postbus 9513, 2300 RA Leiden, The Netherlands
         \and
	         I.Physikalisches Institut, Universit\"at zu K\"oln, 
    	     Z\"ulpicher Str. 77, 50937 K\"oln, Germany}

\authorrunning{N. L\"utzgendorf}

\titlerunning{Intermediate-mass black holes in Globular Clusters}

\abstract{
For a sample of nine Galactic globular clusters we measured the inner kinematic profiles with integral-field spectroscopy that we combined with existing outer kinematic measurements and HST luminosity profiles. With this information we are able to detect the crucial rise in the velocity-dispersion profile which indicates the presence of a central black hole. In addition, N-body simulations compared to our data will give us a deeper insight in the properties of clusters with black holes and stronger selection criteria for further studies. For the first time, we obtain a homogeneous sample of globular cluster integral-field spectroscopy which allows a direct comparison between clusters with and without an intermediate-mass black hole. 
}

\maketitle{}

\section{Introduction}

Intermediate-mass black holes  (IMBHs, $M_{\bullet} \sim 10^2 - 10^5 \, M_{\odot}$)  have drawn the attention of astronomers for more than a decade and would shed light onto the mystery of the rapid growth of supermassive black holes by acting as seeds in the early stage of galaxy formation. 
Besides X-ray and radio emissions, the central kinematics in globular clusters can reveal possible IMBHs. However, resolving the gravitational sphere of influence for plausible IMBH masses ($1'' - 2 ''$ for large Galactic globular clusters) requires velocity-dispersion measurements at a high spatial resolution. Today, with the existence of the Hubble Space Telescope (HST) and high spatial resolution ground based integral-field spectrographs, the search for IMBHs is possible.

A strong motivation for searching for intermediate-mass black holes is the observed relation between the black-hole mass and the velocity dispersion of its host galaxy \citep[e.g.][]{ferrarese_2000, gebhardt_2000b, gultekin_2009}. Exploring this relation in the lower mass range, where we find intermediate-mass black holes, will give important information about the origin and validation of this correlation.

The formation of intermediate-mass black holes can occur by the direct collapse of very massive first generation stars \citep[Population III stars,][]{madau_2001}, or runaway merging in dense young star clusters \citep[e.g.][]{zwart_2004}. This makes globular clusters excellent environments for intermediate-mass black holes. The currently best candidates for hosting an intermediate-mass black hole at their centers are the globular clusters $\omega$ Centauri \citep[NGC 5139,][]{noyola_2008, noyola_2010,vdMA_2010,jalali_2012}, G1 in M31 \citep[][]{gebhardt_2005}, and NGC 6388 \citep{nora11}. All of these very massive globular clusters show kinematic signatures of a central dark mass in their velocity-dispersion profiles. For $\omega$ Centauri, \cite{vdMA_2010} investigated a large dataset of HST proper motions and found less evidence for a central IMBH than proposed by \cite{noyola_2008}. Using a new kinematic center, however, \cite{noyola_2010} and \cite{jalali_2012} confirmed the signature of a dark mass in the center of $\omega$ Centauri and proposed a $\sim 40~000 \, M_{\odot}$ IMBH.

Further evidence for the existence of IMBHs is the discovery of ultra luminous X-ray sources at non-nuclear locations in starburst galaxies \citep[e.g.][]{matsumoto_2001, fabiano_2001}. The brightest of these compact objects (with $L \sim 10^{41} \ \mathrm{erg}\, \mathrm{s}^{-1}$) imply masses larger than $10^3 M_{\odot}$ assuming accretion at the Eddington limit. 

\section{Observations}

Our sample consists of nine Galactic globular clusters, each of them a good candidate for hosting an IMBHs at its center due its high mass and central cusp in its light profile. 

These clusters were observed with the GIRAFFE spectrograph of the FLAMES (Fiber Large Array Multi Element Spectrograph) instrument at the Very Large Telescope (VLT) using the ARGUS mode (Large Integral Field Unit). The velocity-dispersion profile was obtained by combining the spectra in radial bins centered around the adopted photometric center and measuring the broadening of the lines using a non parametric line-of-sight-velocity-distribution fitting algorithm \citep[for more detailed description see][and Feldmeier et al. 2013, in prep.]{nora11,nora12a,nora12c}.  For larger radii, the kinematic profiles were completed with radial velocity from the literature, if existent. In addition to the spectroscopic data, HST photometry was used to obtain the star catalogs, the photometric center of the cluster and its surface brightness profile. For each cluster both photometry and spectroscopy were combined in order to apply analytic Jeans models to the data. The surface brightness profile was used to obtain a model velocity-dispersion profile which was fit to the data by applying different black-hole masses and $M/L_V$ profiles. The final black-hole masses were obtained from a $\chi^2$ fit to the kinematic data. 

\begin{figure*}[]
\resizebox{\hsize}{!}{\includegraphics[clip=true]{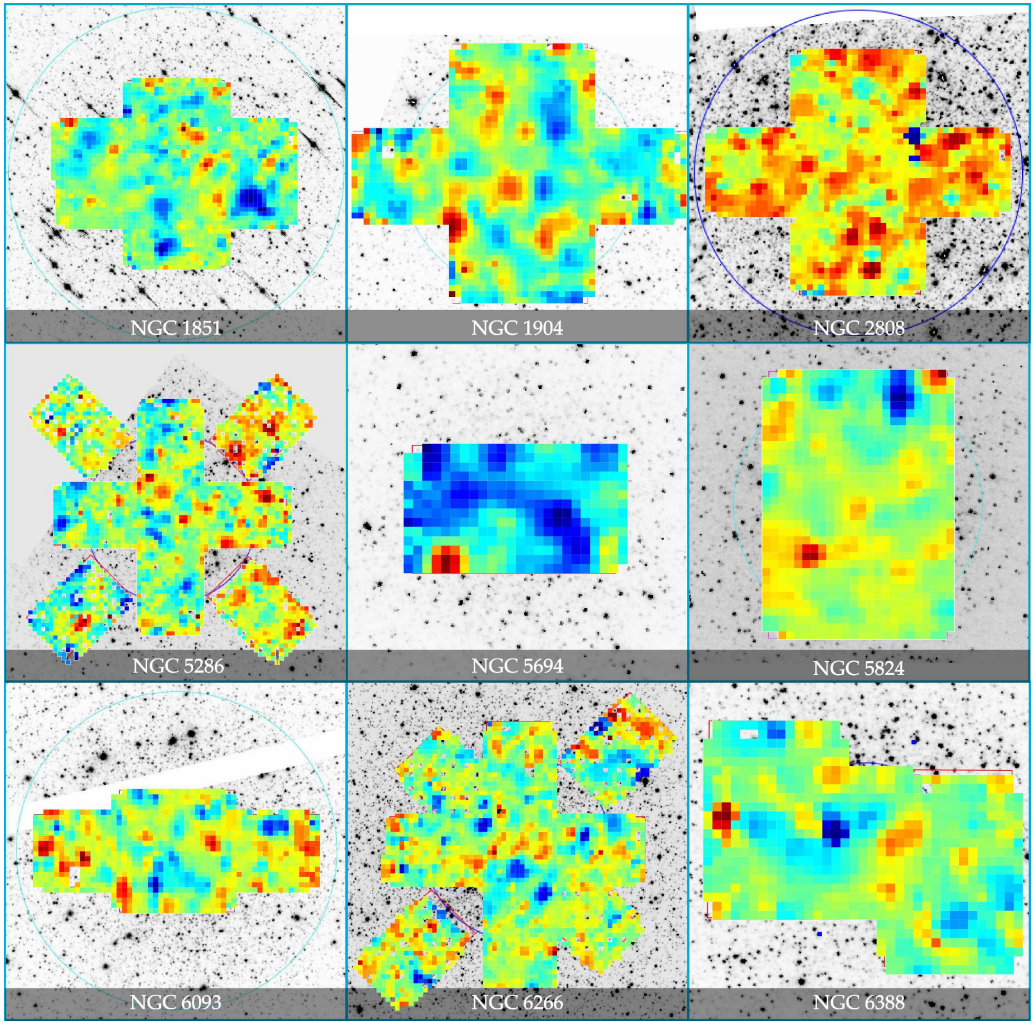}}
\caption{
\footnotesize
Velocity maps of the nine Galactic globular clusters of our sample.
}
\label{vmap}
\end{figure*}

\section{N-body simulations}

Furthermore, we run N-body simulations based on the GPU (Graphic Processing Unit) - enabled version of the collisional N-body code NBODY6 \citep{aarseth_1999}. The code treats binary interactions, stellar evolution, and external tidal fields in a highly sophisticated manner which allows detailed simulations of globular clusters with and without central IMBHs and their observable effects. The goals of these simulations can be summarized in three major points: 

\textit{1) Reproduce Observations:} Our observed clusters will be compared to a grid of scaled N-body simulations in order to find the best fitting initial parameters and constrain the black-hole mass, $M/L_V$ profiles and mass functions. In contrast to Jeans models, N-body simulations can consider the effect of bright stars on the velocity measurements as well as strong mass segregation of stellar remnants in the center. \textit{2) Test our Analysis:} With the output of the N-body simulations we are creating mock datasets of HST images and IFU datacubes in order to test our analyzing methods in photometry and spectroscopy. This will give new insights on how reliable the method of integrated light is in terms of recovering the inner dynamics of dense stellar systems. \textit{3) Study cluster properties:} We are analyzing cluster simulations with different black-hole retention fractions, IMBH masses and binary fractions in tidal fields. With these models we will be able to put constraints on our preliminary black-hole masses, verify the measurements and predict new observables for finding IMBHs.

\begin{figure}[]
\resizebox{\hsize}{!}{\includegraphics[clip=true]{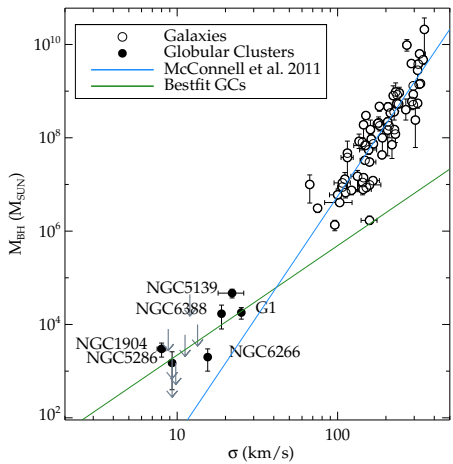}}
\caption{
\footnotesize
$M_{\bullet} - \sigma$ relations of IMBHs and SMBHs in comparison. The slope of the bestfit to the GCs (green line) is by a factor of two smaller than the slope of the SMBHs in galaxies (blue line).
}
\label{msig}
\end{figure}

\section{Results and Conclusions}

We have investigated the presence of IMBHs in nine Galactic globular clusters using a combination of HST photometry and integral-field spectroscopy. Comparing the velocity-dispersion profiles in the central region of each GCs with analytical Jeans models yielded four candidates which show an IMBH signature in their kinematics \citep[see][and Feldmeier et al. 2013, in prep.]{nora11,nora12a,nora12c}. Using our results and IMBH measurements from the literature, we study the $M_{\bullet} - \sigma$ relation for IMBHs and compare it to supermassive black holes in galaxies. Figure \ref{msig} (L\"utzgendorf et al. 2013a, in prep.) shows this comparison and indicates a lower slope for the correlation at the lower mass end compared to the slope obtained for the SMBHs by \citet{McConnell_2011}. The reason for this behavior could be explained by high mass loss of the globular clusters due to tidal stripping. This could lower the velocity dispersion of the system and moves the globular clusters away from the $M_{\bullet} - \sigma$ relation.

In addition to the observational work we performed N-body simulations in order to address open question and degeneracy. The simulations include stellar evolution, binary formation and an external tidal field while the IMBH mass, the black-hole retention fraction and the primordial binary fraction are varied. With the outcome of these simulations we will be able to reproduce our observations, test our analysis methods, and get a deeper understanding of the observable effects of IMBHs to their environment (L\"utzgendorf et al. 2013b, in prep.).



\bibliographystyle{aa}

\end{document}